\documentclass[conference]{IEEEtran}
\IEEEoverridecommandlockouts
\usepackage{cite}
\usepackage{amsmath,amssymb,amsfonts}
\usepackage{algorithmic,algorithm}
\usepackage{graphicx}
\usepackage{textcomp}
\usepackage{xcolor}
\usepackage{tikz}
\usepackage{pgfplots}
\usepackage{pgfplotstable}
\usepackage{bm}
\usepackage{caption}
\usepackage{subcaption}

\def  \codebook{\Phi}

\def\thard{ t^{b,*}_i }

\def\mod{\text{mod}}

\def\C{\mathbb{C}}
\def\ChannelH{H}

\def\BibTeX{{\rm B\kern-.05em{\sc i\kern-.025em b}\kern-.08em
    T\kern-.1667em\lower.7ex\hbox{E}\kern-.125emX}}

\begin{document}

\def\P{ \mathbb{P} }

\def\Hmin{ H_{\min} }
\def\Cmin{ C_{\min} }
\newcommand{\krd}[1]{\textcolor{violet}{ [KRD: #1 ] \normalsize }}
\newcommand{\mg}[1]{\textcolor{blue}{ [MG: #1 ] \normalsize }}

\def\LR{\text{LR}}
\def\LLR{\text{LLR}}
\def\rLLR{L}  
\newcommand{\absLLR}[1]{|\LLR(Y_{#1})|}
\newcommand{\absorbLLR}[1]{L_{#1}}

\title{
Using channel correlation to improve decoding -- ORBGRAND-AI\\
}

\author{
\IEEEauthorblockN{Ken R. Duffy}
\IEEEauthorblockA{\textit{Dept. of ECE \& Dept. Mathematics} \\
\textit{Northeastern University}\\
Boston, USA \\
k.duffy@northeastern.edu}
\and
\IEEEauthorblockN{Moritz Grundei}
\IEEEauthorblockA{\textit{Electrical and Computer Engineering} \\
\textit{Technical University of Munich}\\
Munich, Germany \\
moritz.grundei@tum.de}
\and
\IEEEauthorblockN{Muriel M\'edard}
\IEEEauthorblockA{\textit{Research Laboratory for Electronics} \\
\textit{Massachusetts Institute of Technology}\\
Cambridge, USA \\
medard@mit.edu}
}

\maketitle

\begin{abstract}
To meet the Ultra Reliable Low Latency Communication (URLLC) needs of modern applications, 
there have been significant advances in the development
of short error correction codes and corresponding soft detection
decoders. A substantial hindrance to delivering low-latency is, however, the reliance on
interleaving to break up omnipresent channel correlations
to ensure that decoder input matches decoder assumptions. Consequently, even when using short codes, the need to wait
to interleave data at the sender and
de-interleave at the receiver results in significant latency that acts contrary to
the goals of URLLC. Moreover, interleaving reduces channel capacity, so that potential decoding performance
is degraded.

Here we introduce a variant of Ordered Reliability Bits 
Guessing Random Additive Noise Decoding (ORBGRAND), which we call ORBGRAND-Approximate Independence (ORBGRAND-AI),
a soft-detection decoder that can decode any moderate redundancy code and overcomes the limitation 
of existing decoding paradigms by leveraging channel correlations and circumventing
the need for interleaving. By leveraging correlation,
not only is latency reduced, but error correction 
performance can be enhanced by multiple dB, while decoding complexity is also reduced, offering one potential
solution for the provision of URLLC.

\end{abstract}


\section{Introduction}

Modern applications, including augmented and virtual reality,
vehicle-to-vehicle communications, the Internet of Things, and
machine-type communications, have driven demand for Ultra-Reliable
Low-Latency Communication (URLLC)
\cite{durisi2016toward,she2017radio,chen2018ultra,parvez2018survey,medard20205}.
As traditionally deployed soft detection error correction codes, such as Turbo Codes 
and Low Density Parity Check Codes, are generally only effective for long blocks of data, 
innovations in soft-detection short code decoding has been necessary to meet the
newly imposed latency constraints.

The most notable success to date has been the development of CRC-Aided Polar (CA-Polar) Codes, 
which result from augmenting Polar codes~\cite{Arikan09} 
with an outer Cyclic Redundancy Check (CRC) code and their code-specific CRC-Assisted Successive Cancellation List (CA-SCL) soft detection decoder, e.g. \cite{niu2012crc,tal2015list,balatsoukas2015llr,liang2016hardware}. The promise of accurate soft detection decoding of CA-Polar codes at short block lengths
has resulted in their adoption for all control channel communications in the 5G New Radio (NR) standard~\cite{3gpp38212}.

For URLLC, low latency is required as well as high reliability. In this paper, we show that actively making use of channel correlation rather than removing it through interleaving, as is commonly done, can help with both aspects simultaneously. It has been known since at least the 1970s that, for example, temporal correlation in wireless channels is ubiquitous, with strength depending on the environment and operating regime \cite{Jakes74}. Interleaving diminishes channel correlation in order to provide white noise to the decoder, but correlated noise has lower entropy \cite{Gal68,Cover91} so that the original, correlated channel had higher capacity. 
Here we realize the possibility afforded by preserving correlation to enable better decoding performance at much higher code-rates than can be achieved with an interleaved channel.


Given that interleavers are theoretically known to be detrimental to rate and reliability, it is worthwhile considering why they are currently used.
Their need arises as soft detection decoding approaches  
typically assume that each bit in a communication is impacted independently by noise, resulting in probabilistically independent per-bit reliabilities \cite{lin2004error}. As real-world noise and interference are subject to temporal correlations that result in correlated reliabilities, in the absence of interleaving the mismatched input to decoders would result in degraded error correction performance. 

With its approach to decoding by identifying noise-effects and inferring code-words, Guessing Random Additive Noise Decoding (GRAND) \cite{duffy19GRAND} is well-positioned to embrace noise correlation and decode without interleaving. We have recently shown that in the hard-detection setting GRAND can exploit statistically described Markovian correlation structures to enhance decoding performance \cite{an20,An22}, but the approach taken there cannot be carried over to the soft detection setting, which requires distinct innovation. 

By adopting sensibilities from thermodynamic probability theory to manage dependence \cite{Sullivan1976,Lewis95B,Pfister02,Pfister04B} and combining them with symbol-level ORBGRAND \cite{An22b}, here we demonstrate that it is possible to accurately soft detection decode any moderate redundancy code without the need for interleaving. By removing the interleaver, decoding performance can be enhanced by multiple dB, while complexity and latency are reduced, offering a potential route forward for delivering URLLC. Our approach uses Approximate Independence (AI) with ORBGRAND (ORBGRAND-AI) to obtain these large gains in decoding performance while leveraging dependence over merely small, manageable neighborhoods of symbols. 

\section{ORBGRAND-AI}

\subsection{ORBGRAND}
ORBGRAND is a soft detection variant of GRAND that generates approximate maximum likelihood (ML) decodings for any moderate redundancy code by availing of probabilistically independent binary reliability information consistent with interleaved output \cite{duffy2021ordered,duffy22ORBGRAND}. It is well-suited to efficient implementation in hardware, eg. \cite{abbas2022high,Riaz23}, and has been proven to be almost capacity achieving~\cite{liu2022orbgrand, Yuan22}. 
After obtaining per-bit reliabilities from a receiver, ORBGRAND rank orders them from least reliable to most reliable. A piece-wise linear approximation to the resulting curve is then used to direct the efficient construction of putative noise effect sequences in decreasing order of likelihood. By taking the demodulated signal and sequentially inverting noise effect sequences in decreasing order of likelihood until a code-word is found, an approximate ML decoding is identified. 

Symbol-level ORBGRAND \cite{An22} introduced a modulation-aware variant that assumes symbols experience independent channel effects consistent with symbol-level interleaving. Given a hard detected symbol, its neighbors in the constellation are considered as potential substitutions. The exceedance distance between potential substitution symbols and the hard detected symbol is used as a reliability input for ORBGRAND's rank ordering, whereupon the original noise effect pattern generator is employed. If a symbol substitution pattern proposes a single symbol be substituted more than once, the pattern is discarded. Empirical results demonstrate that symbol-level ORBGRAND can achieve identical performance to operating on bit level reliabilities while realizing a reduction in pattern generation complexity.

\subsection{ORBGRAND-AI}

To enable a receiver to detect or correct errors, prior to transmission each collection of $k$ information bits is coded to a $n>k$ bit code-word $c^n=(c_1,\ldots,c_n)\in\{0,1\}^n$. For spectral efficiency, most communication systems employ high-order modulation where each transmitted symbol communicates multiple bits of information \cite{Proakis}. If a modulation scheme is employed with a complex constellation of size $|\chi|=2^{m_s}$, the $n$ coded bits are translated into $n_s=n/m_s$ symbols by sequentially mapping each collection of $m_s$ bits to the corresponding higher order symbol. In the absence of interleaving, this results in the transmission of the higher order sequence $\mod(c^n) = x^{n_s} = (x_1,\ldots,x_{n_s}) \in \chi^{n_s}$.

Transmissions are impacted by channel effects and noise that cause the received signal sequence to be perturbed. The complex received vector can be written as 
\begin{align*}
Y^{n_s} = (Y_1,\ldots,Y_{n_s}) = \ChannelH x^{n_s}+N^{n_s}, 
\end{align*}
where we assume that the receiver has perfect Channel State Information (CSI), and so knows both $H\in\C^{n_s\times n_s}$ and possesses a probabilistic description of $N^{n_s}$, e.g. that it is complex-valued white Gaussian noise with known variance.

For ORBGRAND-AI's operation, each received signal corresponding to a coded transmission is split into non-overlapping blocks of $b$ symbols, where for notational ease we assume $n_s/b$ is an integer:
\begin{align*}
Y^{n_s} & = \overbrace{
    (Y_1,\ldots,Y_b \,\vert\, \underbrace{ Y_{b+1},\ldots, Y_{2b}}_\text{$b$ symbols} \,\vert\, \cdots \,\vert\, Y_{n_s-b+1},\ldots, Y_{n_s})
    }^\text{$n_s$ symbols} \\
    & = (Y^b_1,\ldots,Y^b_{n_s/b}).
\end{align*}
Each block $i\in\{1,\ldots,n_s/b\}$ of $b$ symbols, $Y^b_i$, is treated separately, with the likelihoods
\begin{align*}
        p_{X^b_i|Y^b_i}(t^b_i|Y^b_i) \text{ for each } t^b_i\in\chi^b
\end{align*}
being evaluated using the channel model and CSI. We define
\begin{align*}
\thard = \arg\max p_{X^b_i|Y^b_i}(t^b_i|Y^b_i)
\end{align*}
to be the symbol-level hard demodulation of the block $Y^b_i=(Y_{(i-1)b+1},\ldots, Y_{ib})$, which takes channel memory over the block into account. 

The core approximation when evaluating the posterior probability of a noise effect sequence $t^{n_s}\in\chi^{n_s}$ describing symbols to be swapped is that the blocks are assumed to be independent, resulting in the following expression
\begin{align*}
        &p_{X^{n_s}|Y^{n_s}}(t^{n_s}|Y^{n_s})\\
        &= \prod_{i=1}^{n_s/b}p_{X_i^b|Y_i^b}(\thard|Y_i^b)\prod_{i=1}^{n_s/b}\dfrac{p_{X_i^b|Y_i^b}(t_i^b|Y_i^b)}{p_{X_i^b|Y_i^b}(\thard|Y_i^b)},
\end{align*}
which has a common term associated to the sequence of all hard-demodulated blocks and each noise effect sequence that swaps a block experiences a likelihood penalty. 

\begin{algorithm}
\caption{ORBGRAND-AI inputs: The received signal $Y^{n_s}$, abandonment threshold $\tau$, channel statistics $\Psi$ and a codebook membership check function $\codebook$. }
\label{alg:pseudo-code}
\begin{algorithmic}
\STATE {\bf Inputs}: $Y^n$, $\codebook$, $\tau$, $\Psi$
\STATE {\bf Output}: $c^{n,*}$, d
\STATE $d\leftarrow 0$
\STATE $w^\mu\leftarrow$ compute likelihoods for substitution symbol blocks
\WHILE{ $d < \tau$}
    \STATE $d\leftarrow d+1$ 
    \STATE $e^\mu \leftarrow$ next most likely ORBGRAND pattern for $w^\mu$
    \IF {no substitution conflict}
        \STATE $s^{n_s} \leftarrow$ substitute blocks
        \STATE $c^{n,*} \leftarrow$ demodulate $s^{n_s}$
        \IF{$\codebook(c^{n,*}) = 0$}
           \STATE{\bf return} $c^{n,*}$,  $d$
        \ENDIF
    \ENDIF
\ENDWHILE
\STATE{\bf return} FAILURE
\end{algorithmic}
\hrule
\end{algorithm}

With the blocks of symbols, $t^i_b$, now playing the role of individual symbols, this expression is identical to the one used for symbol-level ORBGRAND, and so the ORBGRAND approach can be used to generate putative noise effect patterns, $t^{n_s}$, in approximately decreasing order of likelihood. In particular, the set of all alternative groups, $\{t^b_i\neq \thard: i\in\{1,\ldots,{n_s}/b\}\}$, to the hard demodulated blocks of symbols contains $\mu = (2^{m_s b}-1)n/(bm_s)$ elements and they are provided as input to symbol-level ORBGRAND. For algorithmic clarity, pseudo-code for ORBGRAND-AI can be found in Algorithm \ref{alg:pseudo-code}.

The principle of treating neighbouring blocks as approximately independent random variables originates from considerations in thermodynamic probability theory where stochastic processes are approximated by product measures across boundaries \cite{Sullivan1976,Lewis95B,Pfister02,Pfister04B}. For a Gauss-Markov channel, information theoretic results in Section \ref{sec:capacity} indicate that 
in order to move half-way between the differential entropy rate of the interleaved channel and the differential entropy of the noise with complete correlation, it's sufficient to set $b=2$, suggesting that only small block sizes are necessary to obtain significant performance gains.

\section{Gauss-Markov differential entropy}
\label{sec:capacity}
To get a heuristic understanding as to why significant gains should be possible even with small block sizes, $b$, consider the Gauss-Markov setting, which is a generalization of an Additive White Gaussian Noise (AWGN) channel with Markovian temporal correlations. By Burg's theorem \cite{Burg75}, it has maximum-entropy for given energy and any first order correlation constraint between neighboring symbols, and so plays the analogous, pessimistic role to AWGN for memoryless channels. 

With variance $\sigma^2$ and correlation coefficient $\rho\in(0,1]$, assume that the continuous noise sequence $\{N^n\}$ is a zero-mean real-valued Gaussian with covariance matrix $\bm{C}\in \mathbb{R}^{n\times n}$ having entries $C_{i,j} = \sigma^2\rho^{|i-j|}$. The normalized differential entropy rate of $N^n$ can be calculated as 
\begin{align*}
          & \dfrac{1}{2} \log(2e\pi) + \dfrac{1}{2n}\log(|\bm{C}|)  \nonumber\\
          & = \dfrac{1}{2} \log(2e\pi) + \dfrac{1}{2} \log(\sigma^{2}) + \dfrac{1}{2} \left(1- \dfrac{1}{n}\right) \log(1-\rho^2),
          \label{eq:GMentropy}
\end{align*}
e.g. eq. (8.34) \cite{Cover91}. The final term encapsulates the decrease in entropy that arises from channel correlation as $\log(1-\rho^2)<0$ for $\rho>0$. In a heavily interleaved channel $\rho=0$ and the final term is zero. If the channel was truly independent for each block of $b$ bits, then $C_{i,j}$ would be $0$ for $|i-j|>b$ and the normalized differential entropy rate would instead be
\begin{align*}
          \dfrac{1}{2} \log(2e\pi) + \dfrac{1}{2} \log(\sigma^{2}) + \dfrac{1}{2} \left(1- \dfrac{1}{d}\right) \log(1-\rho^2),
\end{align*}
where the only difference is the multiplier of the final term, which has changed from $(1-1/n)$ to $(1-1/b)$. Thus, in this setting, to get more than half of the reduction in normalized differential entropy, a block-size of $b=2$ suffices, suggesting significant gains should be available with small blocks.

\section{Performance Evaluation}
\label{sec:perfeval}
For performance evaluation, we consider BPSK modulation and additive complex Gauss-Markov noise. For noise-effect pattern generation, we use basic ORBGRAND \cite{duffy2021ordered, duffy22ORBGRAND} in the symbol context \cite{An22b}. For error correcting codes, we use CA-Polar codes as they are the state-of-the-art short code with a dedicated binary soft detection decoder, but also show results for codes for which no dedicated soft detection decoder exists, CRC codes and Random Linear Codes (RLCs).

\begin{figure}[htbp]
    \centering
      \begin{subfigure}[b]{0.48\textwidth}
         \centering
         \includegraphics[width=\textwidth]{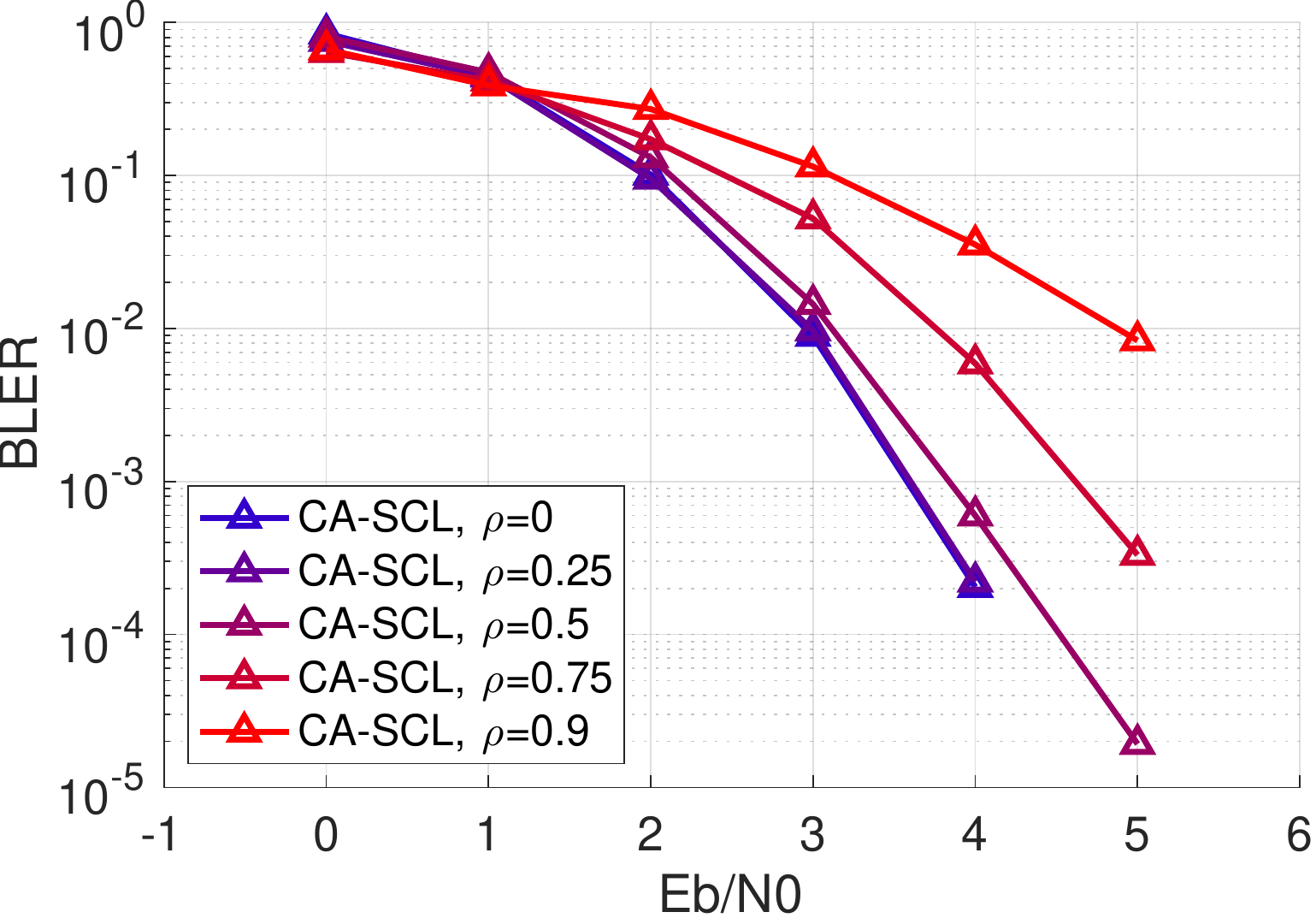}
      \end{subfigure}
    \caption{Impact of channel correlation on CA-SCL decoding of a [128,64] 5G NR CA-Polar code 
    with an 11 bit CRC, list size 8 and within-block interleaver. Block error rate (BLER)
    is plotted versus the energy per information bit, Eb/N0, for a complex Gauss-Markov channel using BPSK modulation with channel correlation strength $\rho$, increasing from blue to red. 
    }
\label{fig:polar}
\end{figure}

We first consider the phenomenon of degraded decoding performance in the presence of channel correlations when the decoder assumes independence. Fig. \ref{fig:polar} plots results for a BPSK modulated 5G NR CA-Polar code that is decoded with its dedicated CA-SCL decoder~\cite{matlab5g}. The within-block CA-Polar code interleaver that is part of the 5G NR standard enables CA-SCL decoding to resist the impact of weak temporal correlation, but performance degrades significantly for moderate to strong correlation owing to the mismatch in assumptions. 

\begin{figure}[htbp]
    \centering
      \begin{subfigure}[b]{0.48\textwidth}
         \centering
         \includegraphics[width=\textwidth]{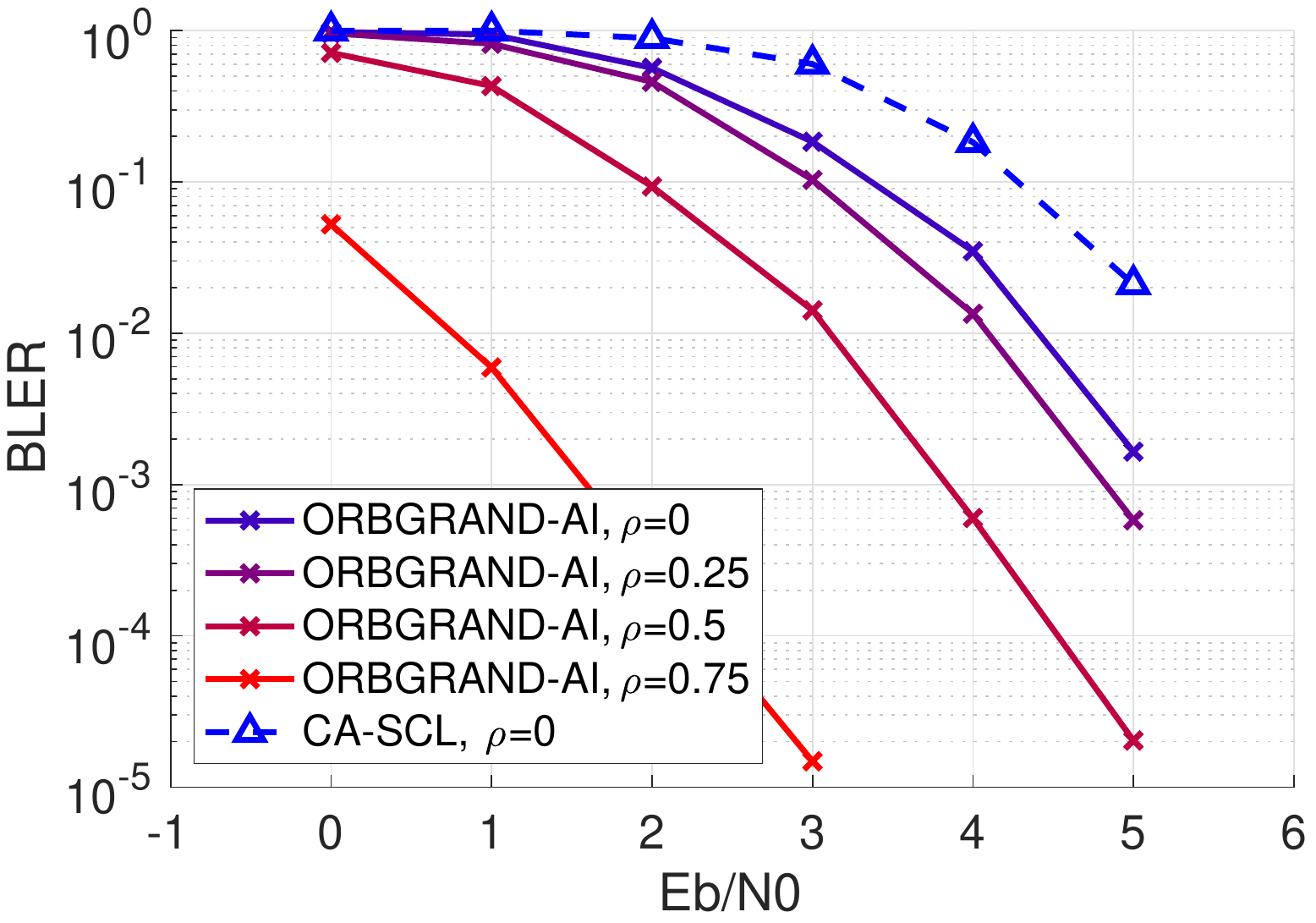}
      \end{subfigure}
    \caption{The impact of channel correlation on ORBGRAND-AI decoding with $b=4$ for a [128,110] 5G NR CA-Polar code with an 11 bit CRC and within-block interleaver for a complex Gauss-Markov channel using BPSK modulation. Channel correlation strength, $\rho$, increases from blue to red. The performance of CA-SCL on an interleaved AWGN channel is shown as a benchmark. 
    }
\label{fig:orbgrand_ai_var_rho}
\end{figure}

Fig. \ref{fig:orbgrand_ai_var_rho} recapitulates the setup in Fig. \ref{fig:polar} 
but with ORBGRAND-AI decoding with a block-size of $b=4$ bits (equivalent to $4$ symbols, as in BPSK each symbol encodes one bit), and for a much higher rate CA-Polar code: $R=0.86$ rather than $R=1/2$. In contrast to CA-SCL decoding, ORBGRAND-AI's BLER performance improves dramatically as channel correlation strength increases. This occurs ORBGRAND-AI exploits the channel correlation to inform its search for the noise effect, resulting in it being identified more accurately. 

In a highly interleaved channel, Fig. \ref{fig:polar} demonstrates that CA-SCL decoding of the [128,64] CA-Polar code provides a BLER of $10^{-3}$ at 3.7dB, which degrades to 4.6dB for $\rho=0.5$. With $\rho=0.5$ ORBGRAND-AI provides the sample BLER performance at the same Eb/N0, but with a code that has only  $18$ redundant bits rather than $64$, and does so without the need for interleaving. With $\rho=0.75$, ORBGRAND-AI achieves that benchmark at 1.5dB, corresponding to a $>$2dB gain over CA-SCL used on the interleaved channel, despite its significantly higher rate and lack of interleaving, demonstrating the substantial potential that comes from exploiting channel correlation to inform decoding.

With BPSK modulation and $b=4$, for a $128$ bit code ORBGRAND-AI considers $32$ blocks where each block has $15=2^4-1$ alternative symbols to the demodulated one. As a result, ORBGRAND-AI sorts $480$ reliability values, which can be readily achieved with a variety of algorithms \cite{abbas2022high,Riaz23}. ORBGRAND's pattern generator on $480$ bit strings can then be efficiently generated using the landslide algorithm \cite{duffy22ORBGRAND}, which has proven to be energy efficient, fast and highly parallelizable in hardware \cite{Riaz23}.
 
\begin{figure}[htbp]
    \centering
      \begin{subfigure}[b]{0.48\textwidth}
         \centering
         \includegraphics[width=\textwidth]{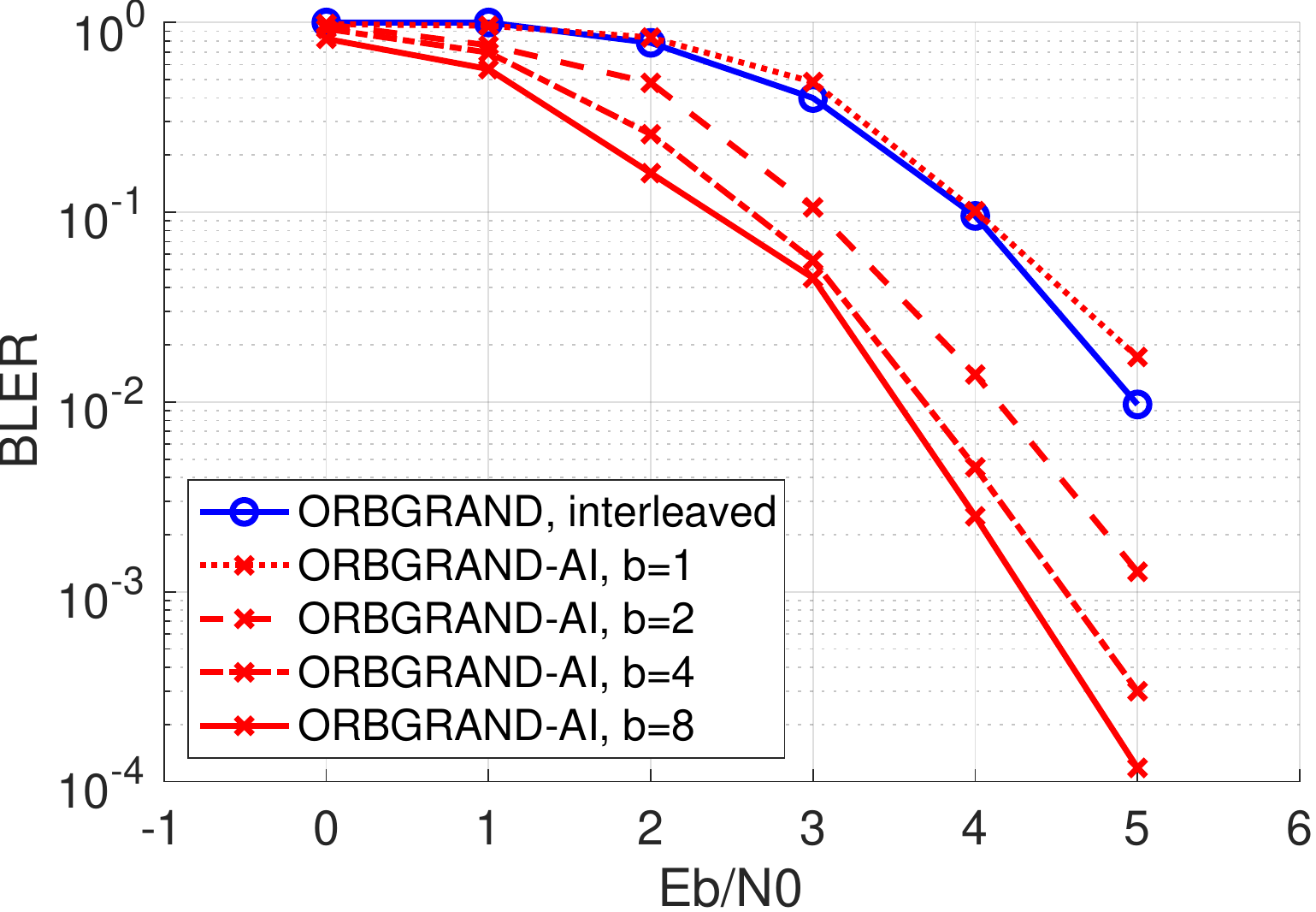}
      \end{subfigure}
    \caption{The impact of block-size, $b$, on BLER performance of ORBGRAND-AI for a fixed $\rho=0.5$ and an [128,116] RLC.}
\label{fig:orbgrand_ai_var_b}
\end{figure}

Fig. \ref{fig:orbgrand_ai_var_b} illustrates the impact of the block-size, $b$, on performance for a Gauss-Markov channel with correlation coefficient $\rho=0.5$. As anticipated by theory, substantial gains are obtained even with small $b$ values and there is a law of diminishing returns as $b$ increases. While $b=8$ provides increased mildly performance, $b=4$ suffices to get substantial gains. When $b=1$, the decoder is assuming that channel effects are independent, even though they are not, and a mild degredation in performance is experienced compared to a fully interleaved system, which is in-line with observations of Fig. \ref{fig:polar} for CA-SCL.

\begin{figure}[htbp]
    \centering
      \begin{subfigure}[b]{0.48\textwidth}
         \centering
         \includegraphics[width=\textwidth]{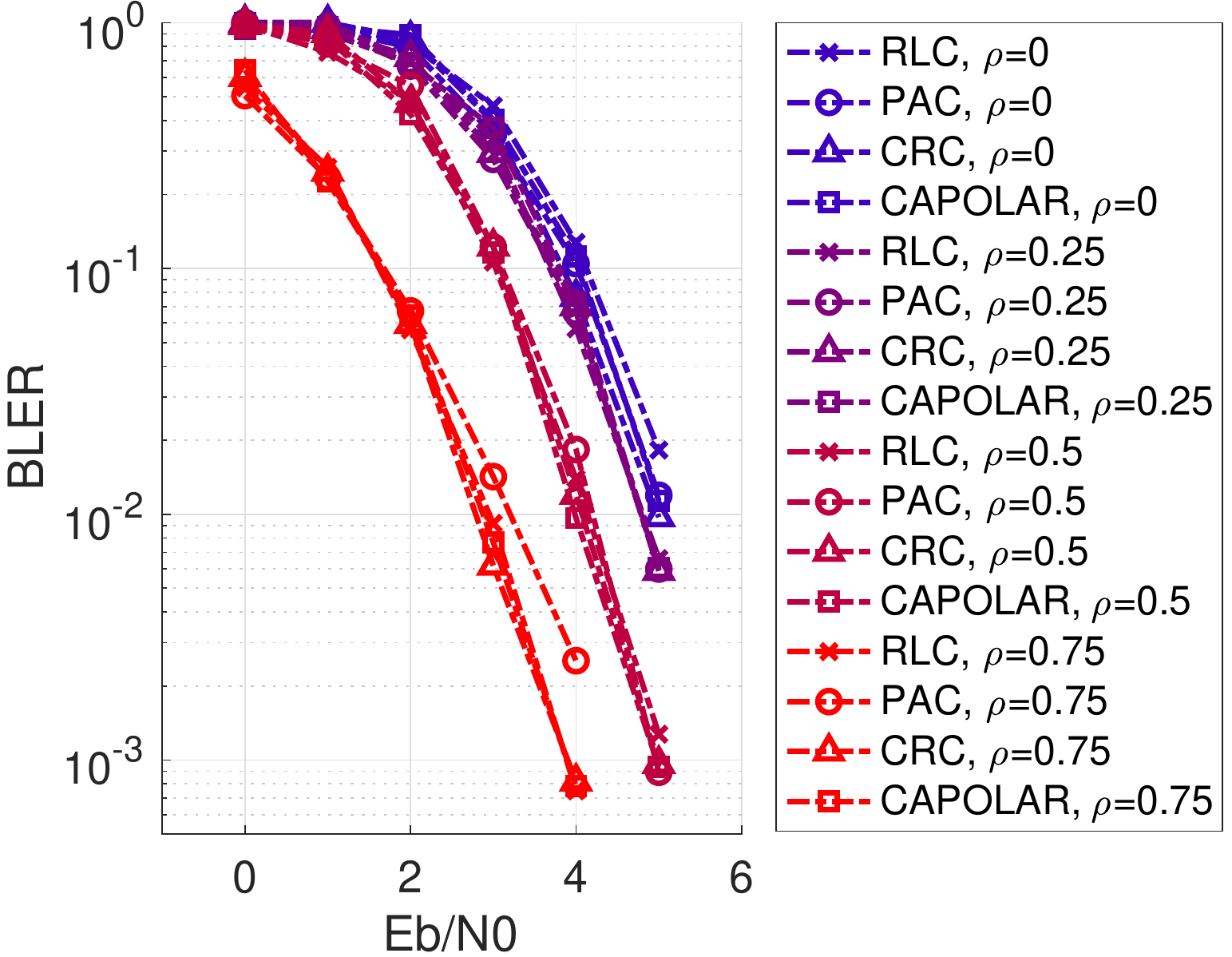}
      \end{subfigure}
    \caption{With $b=2$, ORBGRAND-AI's performance with $b=2$ in a complex Gauss-Markov channel. Shown are results for [128, 116] CA-Polar, RLC, CRC and PAC codes.}
\label{fig:orbgrand_ai_var_code}
\end{figure}

As is known in theory for hard detection decoding \cite{Coffey90}, previous studies with GRAND and ORBGRAND have consistently shown that nearly all code structures of the same dimensions provide similar error correction performance, e.g. \cite{Riaz21,Papadopoulou21,An22,duffy22ORBGRAND}, even non-linear ones \cite{cohen22}. Fig. \ref{fig:orbgrand_ai_var_code}
demonstrates that this phenomenon extends to decoding while exploiting channel correlation. Results are shown for CA-Polar codes, RLCs, CRC codes and Polar-Assisted Convolutional codes \cite{Arikan19PAC}, which were recently introduced by the originator of Polar codes. For each value of $\rho$, it can be seen that all four code types provide similar decoding performance and so, in general, no particular code-structure is preferred.

\begin{figure}[htbp]
    \centering
      \begin{subfigure}[b]{0.48\textwidth}
         \centering
         \includegraphics[width=\textwidth]{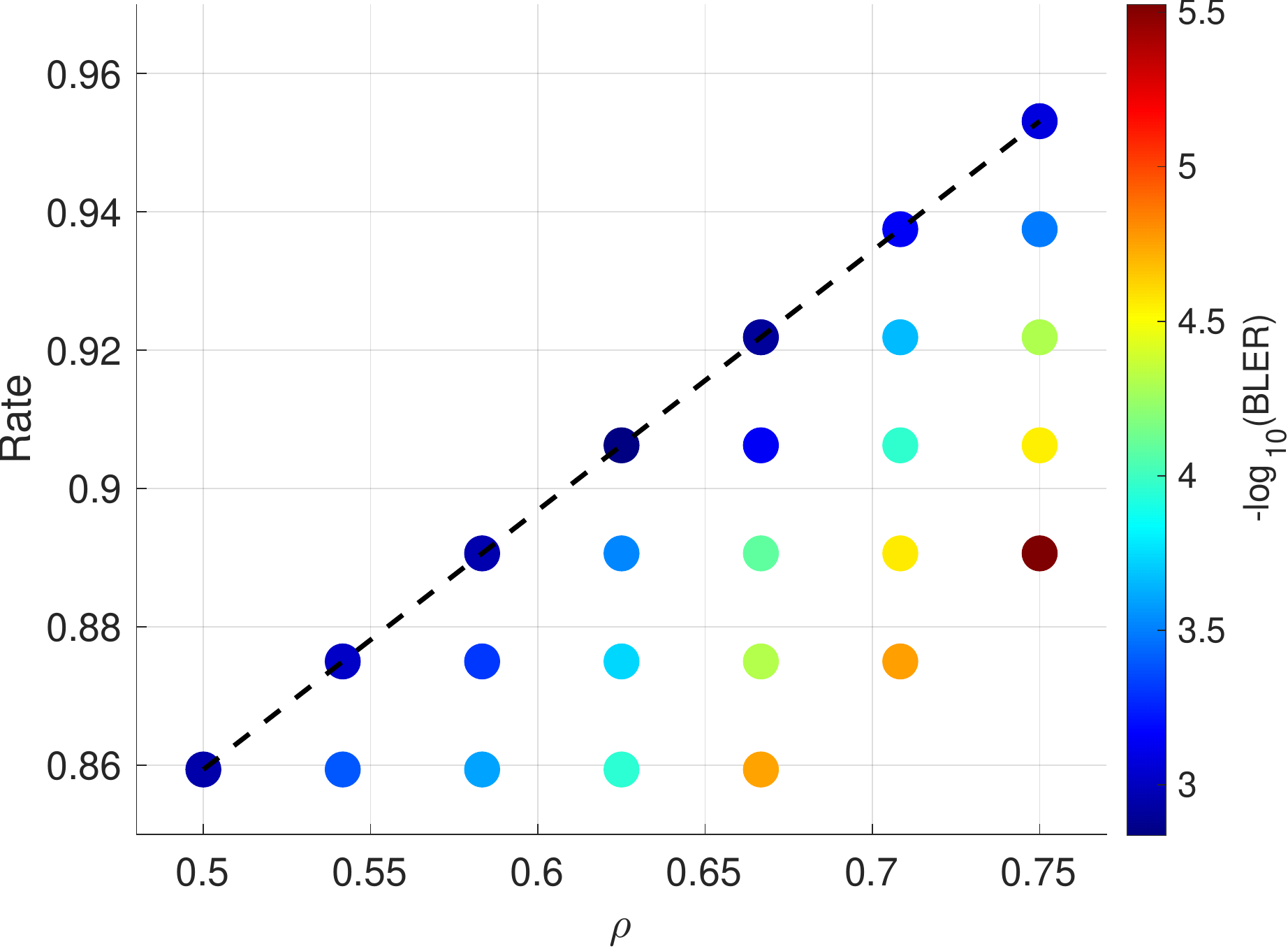}
      \end{subfigure}
    \caption{Code-rate versus $\rho$ to achieve a BLER of no more than $10^{-3}$ at 3.7dB when RLCs of length 128 are decoded with ORBGRAND-AI and a block size of $b=4$.}
\label{fig:orbgrand_ai_var_R}
\end{figure}

Fig. \ref{fig:orbgrand_ai_var_R} further explores the phenomenon exhibited in Fig. \ref{fig:orbgrand_ai_var_rho} where better BLER performance than a state-of-the-art rate $1/2$ 5G NR CA-Polar code decoded with CA-SCL can be achieved with substantially higher rate codes in the presence of correlation by using ORBGRAND-AI. With a target benchmark of obtaining a BLER of $10^{-3}$ or lower at Eb/N0=3.7dB, the heat-map shows the BLER performance of ORBGRAND-AI decoding of RLCs as a function a matrix of channel-correlation strength and code-rate. As $\rho$ increases from $0.5$ to $0.75$, while performing as well as the benchmark of the rate 1/2 interleaved code, it is possible to approximately linearly increase the code rate from $0.86$ to $0.95$ without interleaving while achieving decoding precision as good as CA-SCL obtains with a rate $1/2$ CA-Polar code on a fully interleaved channel. 

\begin{figure}[htbp]
    \centering
      \begin{subfigure}[b]{0.48\textwidth}
         \centering
         \includegraphics[width=\textwidth]{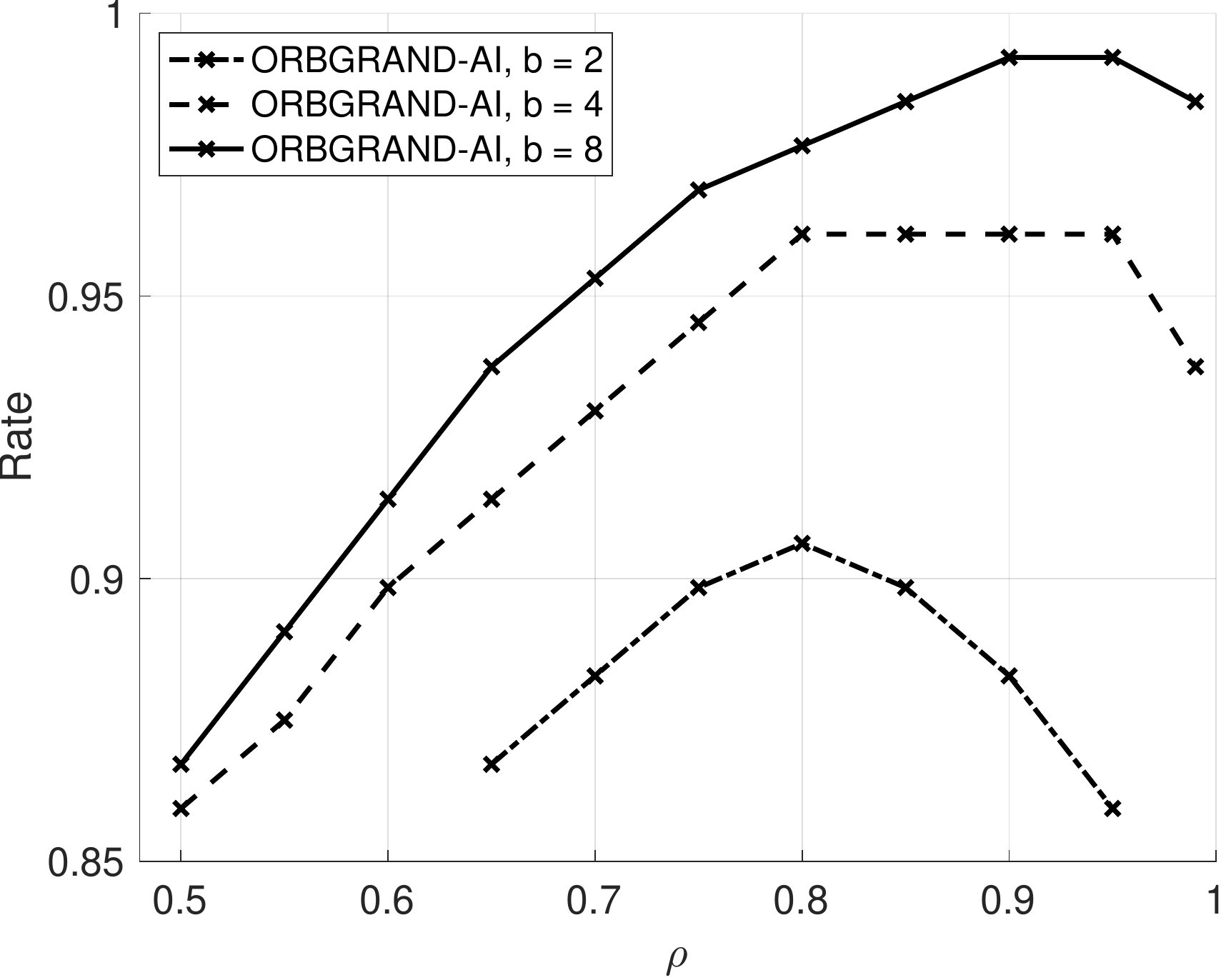}
      \end{subfigure}
    \caption{Code-rate versus $\rho$ to achieve a BLER of $10^{-3}$ at 3.7dB when RLCs of length 128 are decoded with ORBGRAND-AI varying block size $b$.}
\label{fig:orbgrand_ai_var_R_var_b}
\end{figure}

Complementing Fig. \ref{fig:orbgrand_ai_var_R}, Fig. \ref{fig:orbgrand_ai_var_R_var_b} presents the highest rate RLC for which ORBGRAND-AI provides a BLER of less than $10^{-3}$ at Eb/N0=3.7dB, the benchmark provided by a CA-Polar[128,64] code decoded with CA-SCL on an interleaved channel, for a range of AI-block-sizes, $b$. Higher-rates provide better BLER with larger $b$ as more of the correlation is directly exploited, with diminishing returns moving from $2$ to $4$ to $8$. While significant gains are seen for all values, as correlation increases eventually the smaller block-size cannot fully exploit the benefit.

\section{Discussion}
In order to deliver low latency communications, significant progress has been made in the development of short error correction codes with dedicated soft-detection decoders, notably CA-Polar codes and CA-SCL decoding. The use of interleavers ensures that channel effects meet decoder assumptions of independent per-bit reliabilities, but hinders attempts to provide low-latency communication. Here we demonstrate that ORBRGAND-AI can drop interleavers and exploit channel correlation to dramatically improve decoding performance. As a result, similar BLER performance to a rate-$1/2$ CA-Polar code can be obtained with a much higher rate code and with no added latency from interleaving. Moreover, as ORBGRAND-AI can efficiently decode any moderate redundancy code, this performance can be achieved without the need to develop and deploy bespoke codes. Indeed, CRCs, which are widely deployed and traditionally only used for error detection, perform as well as CA-Polar codes.

While an application of the data processing inequality tells us that the Shannon capacity of channels with memory can only be higher than those without \cite{Cover91}, which indicates that the optimal decoding performance available on interleaving channels must be lesser than on non-interleaved channels, the extent to which leveraging correlation in decoding improves performance may be surprising. Considerations of differential noise entropy, however, suggests that making use of the joint dependence of even small collections of symbols should provide substantially improved performance, and this holds true to simulation results.

We have illustrated the approach here for Gauss-Markov channels, which arise naturally when noise is additive white Gaussian, CSI is known and channel effects are inverted. As there is an equivalence between channels with colored Gaussian noise and intersymbol-interference (ISI) channels \cite{Tsy70, Gray72,Cover91}, these results directly apply to that setting. The approach, however, can be used more generally for any system with CSI as the channel descriptor's role is merely in evaluating the likelihoods of alternative blocks of symbols. Indeed, the method could be used to improve the performance of MIMO decoding, for example. If CSI is imperfect, previous work has shown GRAND algorithms can be used to improve its estimation \cite{Sarieddeen22, Sarieddeen22soft}, which could also be leveraged in the context of ORBGRAND-AI.

As ORBGRAND's pattern generator has been proven to be highly energy efficient in hardware, \cite{Riaz23}, ORBGRAND-AI provides one potential way forward in the provision of URLLC without interleaving, but several further opportunities present themselves too. During the execution of any GRAND-algorithm, it has recently been shown that it is possible to evaluate an a posteriori probability that the decoding is correct using simple calculations \cite{duffy22LR}. That information can be exploited to upgrade ubiquitous CRCs from error detection to soft error correction and reduce costly hybrid ARQ requests when decoding is confident. Such confidence measures can be used with ORBGRAND-AI to further utilize widely deployed CRCs. As another example, long, low-rate codes, such as product codes, can be constructed from shorter, higher-rate component codes and the use of GRAND \cite{Galligan21} and ORBGRAND \cite{condo2022_iterative,Galligan23} in their decoding has previously been explored. In the presence of correlation, component decoding performance with would be significantly improved with ORBGRAND-AI to an extent that may outperform Low Density Parity Check (LDPC) codes, which are the long-standing long-code incumbent. 

Note that while we have used the basic ORBGRAND pattern generator that has been implemented in hardware \cite{Riaz23} for the results shown here, there is no difficulty in using the more sophisticated statistical model that provides enhanced results at higher SNR \cite{duffy22ORBGRAND}.

Acknowledgment: This work was supported by Defense Advanced Research Projects Agency contract number HR00112120008.
\bibliographystyle{IEEEtran}
\bibliography{grand,ORBGRAND, moritz}

\begin{thebibliography}{10}
\providecommand{\url}[1]{#1}
\csname url@samestyle\endcsname
\providecommand{\newblock}{\relax}
\providecommand{\bibinfo}[2]{#2}
\providecommand{\BIBentrySTDinterwordspacing}{\spaceskip=0pt\relax}
\providecommand{\BIBentryALTinterwordstretchfactor}{4}
\providecommand{\BIBentryALTinterwordspacing}{\spaceskip=\fontdimen2\font plus
\BIBentryALTinterwordstretchfactor\fontdimen3\font minus
  \fontdimen4\font\relax}
\providecommand{\BIBforeignlanguage}[2]{{%
\expandafter\ifx\csname l@#1\endcsname\relax
\typeout{** WARNING: IEEEtran.bst: No hyphenation pattern has been}%
\typeout{** loaded for the language `#1'. Using the pattern for}%
\typeout{** the default language instead.}%
\else
\language=\csname l@#1\endcsname
\fi
#2}}
\providecommand{\BIBdecl}{\relax}
\BIBdecl

\bibitem{durisi2016toward}
G.~Durisi, T.~Koch, and P.~Popovski, ``Toward massive, ultrareliable, and
  low-latency wireless communication with short packets,'' \emph{Proc. IEEE},
  vol. 104, no.~9, pp. 1711--1726, 2016.

\bibitem{she2017radio}
C.~She, C.~Yang, and T.~Q. Quek, ``Radio resource management for ultra-reliable
  and low-latency communications,'' \emph{IEEE Commun. Mag.}, vol.~55, no.~6,
  pp. 72--78, 2017.

\bibitem{chen2018ultra}
H.~Chen, R.~Abbas, P.~Cheng, M.~Shirvanimoghaddam, W.~Hardjawana, W.~Bao,
  Y.~Li, and B.~Vucetic, ``Ultra-reliable low latency cellular networks: Use
  cases, challenges and approaches,'' \emph{IEEE Commun. Mag.}, vol.~56,
  no.~12, 2018.

\bibitem{parvez2018survey}
I.~Parvez, A.~Rahmati, I.~Guvenc, A.~I. Sarwat, and H.~Dai, ``A survey on low
  latency towards {5G: RAN}, core network and caching solutions,'' \emph{IEEE
  Commun. Surv.}, vol.~20, no.~4, pp. 3098--3130, 2018.

\bibitem{medard20205}
M.~M{\'e}dard, ``Is 5 just what comes after 4?'' \emph{Nature Electronics},
  vol.~3, no.~1, pp. 2--4, 2020.

\bibitem{Arikan09}
E.~{Arikan}, ``Channel polarization: A method for constructing
  capacity-achieving codes for symmetric binary-input memoryless channels,''
  \emph{IEEE Trans. Inf. Theory}, vol.~55, no.~7, pp. 3051--3073, 2009.

\bibitem{niu2012crc}
K.~Niu and K.~Chen, ``{CRC}-aided decoding of {P}olar codes,'' \emph{IEEE
  Commun. Letters}, vol.~16, no.~10, pp. 1668--1671, 2012.

\bibitem{tal2015list}
I.~Tal and A.~Vardy, ``List decoding of {P}olar codes,'' \emph{IEEE Trans. Inf.
  Theory}, vol.~61, no.~5, pp. 2213--2226, 2015.

\bibitem{balatsoukas2015llr}
A.~Balatsoukas-Stimming, M.~B. Parizi, and A.~Burg, ``{LLR}-based successive
  cancellation list decoding of {P}olar codes,'' \emph{IEEE Trans. Signal
  Process.}, vol.~63, no.~19, pp. 5165--5179, 2015.

\bibitem{liang2016hardware}
X.~Liang, J.~Yang, C.~Zhang, W.~Song, and X.~You, ``Hardware efficient and
  low-latency {CA-SCL} decoder based on distributed sorting,'' in \emph{IEEE
  GLOBECOM}, 2016.

\bibitem{3gpp38212}
``{3rd Generation Partnership Project; Technical Specification Group Radio
  Access Network; NR; Multiplexing and Channel Coding, Release 15, V15.6.0},''
  3GPP, 38.212, Tech. Rep., June 2019.

\bibitem{Jakes74}
W.~C. Jakes, \emph{Mobile Radio Propagation}, 1974.

\bibitem{Gal68}
R.~G. Gallager, \emph{Information Theory and Reliable Communication}.\hskip 1em
  plus 0.5em minus 0.4em\relax New York, NY, USA: John Wiley \& Sons, Inc.,
  1968.

\bibitem{Cover91}
T.~M. Cover and J.~A. Thomas, \emph{Elements of Information Theory}.\hskip 1em
  plus 0.5em minus 0.4em\relax John Wiley \& Sons, 1991.

\bibitem{lin2004error}
S.~Lin and D.~J. Costello, \emph{Error control coding: fundamentals and
  applications}.\hskip 1em plus 0.5em minus 0.4em\relax Pearson/Prentice Hall,
  2004.

\bibitem{duffy19GRAND}
K.~R. {Duffy}, J.~{Li}, and M.~{M\'edard}, ``Capacity-achieving guessing random
  additive noise decoding,'' \emph{IEEE Trans. Inf. Theory}, vol.~65, no.~7,
  pp. 4023--4040, 2019.

\bibitem{an20}
W.~An, M.~M{\'e}dard, and K.~R. Duffy, ``Keep the bursts and ditch the
  interleavers,'' in \emph{IEEE GLOBECOM}, 2020.

\bibitem{An22}
W.~An, M.~M\'edard, and K.~R. Duffy, ``Keep the bursts and ditch the
  interleavers,'' \emph{IEEE Trans. Commun.}, vol.~70, pp. 3655--3667, 2022.

\bibitem{Sullivan1976}
W.~G. Sullivan, ``Specific information gain for interacting markov processes,''
  \emph{Probab. Theory Relat. Fields}, vol.~37, no.~1, pp. 77--90, 1976.

\bibitem{Lewis95B}
J.~T. Lewis, C.-E. Pfister, and W.~G. Sullivan, ``Entropy, concentration of
  probability and conditional limit theorems,'' \emph{Markov Process. Relat.
  Fields}, vol.~1, pp. 319--386, 1995.

\bibitem{Pfister02}
C.-E. Pfister, ``Thermodynamical aspects of classical lattice systems,''
  \emph{Prog. Probab.}, pp. 393--472, 2002.

\bibitem{Pfister04B}
C.-E. Pfister and W.~G. Sullivan, ``Large deviations estimates for dynamical
  systems without the specification property. application to the
  $\beta$-shifts,'' \emph{Nonlinearity}, vol.~18, no.~1, p. 237, 2004.

\bibitem{An22b}
W.~An, M.~Medard, and K.~R. Duffy, ``Soft decoding without soft demapping with
  {ORBGRAND},'' \emph{arXiv:2207.11991}, 2022.

\bibitem{duffy2021ordered}
K.~R. Duffy, ``Ordered reliability bits guessing random additive noise
  decoding,'' in \emph{IEEE ICASSP}, 2021, pp. 8268--8272.

\bibitem{duffy22ORBGRAND}
K.~R. Duffy, W.~An, and M.~M{\'e}dard, ``Ordered reliability bits guessing
  random additive noise decoding,'' \emph{IEEE Trans. Signal Process.},
  vol.~70, pp. 4528--4542, 2022.

\bibitem{abbas2022high}
S.~M. Abbas, T.~Tonnellier, F.~Ercan, M.~Jalaleddine, and W.~J. Gross,
  ``High-throughput and energy-efficient vlsi architecture for ordered
  reliability bits {GRAND},'' \emph{IEEE Trans. Very Large Scale Integr.
  Syst.}, vol.~30, no.~6, pp. 681--693, 2022.

\bibitem{Riaz23}
A.~Riaz, A.~Yasar, F.~Ercan, W.~An, J.~Ngo, K.~Galligan, M.~M\'edard, K.~R.
  Duffy, and R.~T. Yazicigil, ``A sub-0.8p{J}/b 16.3{G}bps/mm$^2$ universal
  soft-detection decoder using {ORBGRAND} in 40nm {CMOS},'' in \emph{IEEE
  ISSCC}, 2023.

\bibitem{liu2022orbgrand}
M.~Liu, Y.~Wei, Z.~Chen, and W.~Zhang, ``{ORBGRAND} is almost
  capacity-achieving,'' \emph{arxiv:2202.06247}, 2022.

\bibitem{Yuan22}
P.~Yuan, K.~R. Duffy, E.~Gabhart, and M.~Médard, ``On the role of quantization
  of soft information in {GRAND},'' \emph{arXiv:2203.13552}, 2022.

\bibitem{Proakis}
J.~G. Proakis, \emph{Digital Communications}.\hskip 1em plus 0.5em minus
  0.4em\relax NY, USA: McGraw-Hill, 2001.

\bibitem{Burg75}
J.~Burg, \emph{Maximum Entropy Spectral Analysis}.\hskip 1em plus 0.5em minus
  0.4em\relax Ph.D. Thesis Stanford University, 1975.

\bibitem{matlab5g}
``{{5G} New Radio Polar Coding.}''
  \url{https://www.mathworks.com/help/5g/gs/polar-coding.html}, [Online;
  accessed 17-September-2023].

\bibitem{Coffey90}
J.~T. Coffey and R.~M. Goodman, ``Any code of which we cannot think is good,''
  \emph{IEEE Trans. Inf. Theory}, vol.~36, no.~6, pp. 1453--1461, 1990.

\bibitem{Riaz21}
A.~Riaz, V.~Bansal, A.~Solomon, W.~An, Q.~Liu, K.~Galligan, K.~R. Duffy,
  M.~M\'edard, and R.~T. Yazicigil, ``Multi-code multi-rate universal maximum
  likelihood decoder using {GRAND},'' in \emph{IEEE ESSCIRC}, 2021, pp.
  239--246.

\bibitem{Papadopoulou21}
V.~Papadopoulou, M.~Hashemipour-Nazari, and A.~Balatsoukas-Stimming, ``Short
  codes with near-{ML} universal decoding: are random codes good enough?'' in
  \emph{IEEE SiPS}, 2021, pp. 94--98.

\bibitem{cohen22}
A.~Cohen, R.~G.~L. D'Oliveira, K.~R. Duffy, J.~Woo, and M.~Médard, ``{AES} as
  error correction,'' \emph{arXiv:2203.12047}, 2022.

\bibitem{Arikan19PAC}
E.~Arikan, ``From sequential decoding to channel polarization and back again,''
  \emph{arXiv:1908.09594}, 2019.

\bibitem{Tsy70}
B.~S. Tsybakov, ``Capacity of a~discrete-time gaussian channel with a filter,''
  \emph{Probl. Peredachi Inf.}, vol.~6, no.~3, pp. 78--82, 1970.

\bibitem{Gray72}
R.~Gray, ``On the asymptotic eigenvalue distribution of toeplitz matrices,''
  \emph{IEEE Trans. Inf. Theory}, vol.~18, no.~6, pp. 725--730, 1972.

\bibitem{Sarieddeen22}
H.~Sarieddeen, M.~M{\'e}dard, and K.~R. Duffy, ``{GRAND} for fading channels
  using pseudo-soft information,'' in \emph{IEEE GLOBECOM}, 2022, pp.
  3502--3507.

\bibitem{Sarieddeen22soft}
------, ``Soft-input, soft-output joint detection and grand,'' in \emph{IEEE
  GLOBECOM}, 2022, pp. 6182--6187.

\bibitem{duffy22LR}
K.~R. Duffy and M.~M\'edard, ``Confident decoding with {GRAND},''
  \emph{arxiv:2212.05309}, 2022.

\bibitem{Galligan21}
K.~Galligan, A.~Solomon, A.~Riaz, R.~T. Yazicigil, M.~M\'edard, and K.~R.
  Duffy, ``{IGRAND}: decode any product code,'' in \emph{IEEE GLOBECOM}, 2021.

\bibitem{condo2022_iterative}
C.~Condo, ``Iterative soft-input soft-output decoding with ordered reliability
  bits {GRAND},'' 2022, arXiv:2207.06691.

\bibitem{Galligan23}
K.~Galligan, M.~M\'edard, and K.~R. Duffy, ``Block turbo decoding with
  {ORBGRAND},'' in \emph{CISS}, 2023.

\end{thebibliography}

\end{document}